\documentclass[nofootinbib,preprint]{revtex4}%
\usepackage{hyperref}
\usepackage{amsmath}
\usepackage{amsfonts}
\usepackage{amssymb}
\usepackage{graphicx}%
\providecommand{\U}[1]{\protect\rule{.1in}{.1in}}

\begin{document}
\title{Regge Closed String Scattering and its Implication on Fixed angle Closed
String Scattering }
\author{Jen-Chi Lee}
\email{jcclee@cc.nctu.edu.tw}
\affiliation{Department of Electrophysics, National Chiao-Tung University and Physics
Division, National Center for Theoretical Sciences, Hsinchu, Taiwan, R.O.C.}
\author{Yi Yang}
\email{yiyang@mail.nctu.edu.tw}
\affiliation{Department of Electrophysics, National Chiao-Tung University and Physics
Division, National Center for Theoretical Sciences, Hsinchu, Taiwan, R.O.C.}
\date{\today }

\begin{abstract}
We calculate the complete closed string high energy scattering amplitudes
(HSA) in the Regge regime for arbitrary mass levels. As an application, we
deduce the complete ratios among closed string HSA in the fixed angle regime
by using Stirling number identities. These results are in contrast with the
incomplete set of closed string HSA in the fixed angle regime calculated
previously. The complete forms of the fixed angle amplitudes, and hence the
ratios, were not calculable previously without the input of zero-norm state
calculation. This is mainly due to the lack of saddle point in the fixed angle
closed string calculation.

\end{abstract}
\maketitle

\section{Introduction}

Recently high-energy, fixed angle behavior of string scattering amplitudes
\cite{GM, Gross, GrossManes} was intensively reinvestigated
\cite{ChanLee1,ChanLee2, CHL,CHLTY,PRL,Decay,Compact,susy,CC} for string
states at arbitrary mass levels. The motivation was to uncover the long-sought
hidden stringy spacetime symmetry. A saddle-point method was developed to
calculate the general formula for tree-level high-energy open string
scattering amplitudes of four arbitrary string states. Remarkably, it was
found that there is only one independent component of the amplitudes at each
fixed mass level, and ratios among high energy scattering amplitudes of
different string states at each mass level can be obtained. However, it was
soon realized that \cite{Closed} the saddle-point method was applicable to
$(t,u)$ channel only but not $(s,t)$ channel. It was also pointed out that,
through the observation of the KLT formula \cite{KLT}, this difficulty is
associated with the lack of saddle-point in the integration regime for the
closed string calculation. To calculate the complete high energy closed string
scattering amplitudes in the fixed angle regime \cite{Closed}, one had to rely
on calculation based on the method of decoupling of zero-norm states
\cite{ZNS1,ZNS3,ZNS2} in the spectrum. With this new input, an infinite number
of linear relations among high energy scattering amplitudes of different
string states can be derived and the complete ratios among high energy closed
string scattering amplitudes at each fixed mass level can be determined. One
can now calculate only high energy amplitude corresponding to the highest spin
state at each mass level in the spectrum, and the complete closed string
scattering amplitudes can then be obtained.

In this paper, we will use another method to calculate the closed string
ratios in the fixed angle regime mentioned above. We will calculate the
complete closed string scattering amplitudes in the Regge regime, which have
not been considered in the literature so far. It turned out that both the
saddle-point method and the method of decoupling of zero-norm states adopted
in the calculation of fixed angle regime do not apply to the case of Regge
regime. However a direct calculation is manageable. The calculation will be
based on the KLT formula and the open string $(s,t)$ channel scattering
amplitudes in the Regge regime calculated previously \cite{KLY}. By using a
set of Stirling number identities developed in combinatoric number theory
\cite{MK}, one can then extract the ratios in the fixed angle regime from
Regge closed string scattering amplitudes.

\section{Fixed angle Scattering}

We begin with a brief review of high energy string scatterings in the fixed
angle regime,%
\begin{equation}
s,-t\rightarrow\infty,t/s\approx-\sin^{2}\frac{\theta}{2}=\text{fixed (but
}\theta\neq0\text{)} \label{1}%
\end{equation}
where $s,t$ and $u$ are the Mandelstam variables and $\theta$ is the CM
scattering angle. It was shown \cite{CHLTY,PRL} that for the 26D open bosonic
string the only states that will survive the high-energy limit at mass level
$M_{2}^{2}=2(n-1)$ are of the form%
\begin{equation}
\left\vert n,2m,q\right\rangle \equiv(\alpha_{-1}^{T})^{n-2m-2q}(\alpha
_{-1}^{L})^{2m}(\alpha_{-2}^{L})^{q}|0,k\rangle, \label{2}%
\end{equation}
where the polarizations of the 2nd particle with momentum $k_{2}$ on the
scattering plane were defined to be $e^{P}=\frac{1}{M_{2}}(E_{2}%
,\mathrm{k}_{2},0)=\frac{k_{2}}{M_{2}}$ as the momentum polarization,
$e^{L}=\frac{1}{M_{2}}(\mathrm{k}_{2},E_{2},0)$ the longitudinal polarization
and $e^{T}=(0,0,1)$ the transverse polarization. Note that $e^{P}$ approaches
to $e^{L}$ in the fixed angle regime. For simplicity, we choose $k_{1}$,
$k_{3}$ and $k_{4}$ to be tachyons. It turned out that the $(t,u)$ channel of
the scattering amplitudes can be calculated by using the saddle-point method
and the final results are \cite{CHLTY,PRL,Closed}%
\begin{equation}
\frac{A^{(n,2m,q)}(t,u)}{A^{(n,0,0)}(t,u)}=\left(  -\frac{1}{M_{2}}\right)
^{2m+q}\left(  \frac{1}{2}\right)  ^{m+q}(2m-1)!! \label{3}%
\end{equation}
with%
\begin{align}
A^{(n,0,0)}(t,u)  &  \simeq\sqrt{\pi}(-1)^{n-1}2^{-n}E^{-1-2n}(-2E^{3}%
\sin\theta)^{n}(\sin\frac{\theta}{2})^{-3}(\cos\frac{\theta}{2})^{5-2n}%
\nonumber\\
&  \times\exp(-\frac{t\ln t+u\ln u-(t+u)\ln(t+u)}{2}). \label{4}%
\end{align}
To calculate the high energy, fixed angle closed string scattering amplitudes,
one encountered the well-known difficulty of the lack of saddle-point in the
integration regime. In fact, it was demonstrated \cite{Closed} by three
evidences that the standard saddle-point calculation for high energy closed
string scattering amplitudes was not reliable. It was also pointed out
\cite{Closed} that this difficulty is associated with the lack of saddle-point
in the integration regime for the calculation of $(s,t)$ channel high energy
open string scattering amplitudes. This can be seen from a formula by Kawai,
Lewellen and Tye (KLT), which expresses the relation between tree amplitudes
of closed and open string $(\alpha_{\text{closed}}^{\prime}=4\alpha
_{\text{open}}^{\prime}=2)$ \cite{KLT}
\begin{equation}
A_{\text{closed}}^{\left(  4\right)  }\left(  s,t,u\right)  =\sin\left(  \pi
k_{2}\cdot k_{3}\right)  A_{\text{open}}^{\left(  4\right)  }\left(
s,t\right)  \bar{A}_{\text{open}}^{\left(  4\right)  }\left(  t,u\right)  .
\label{5}%
\end{equation}
Note that Eq.(\ref{5}) is valid for all energies. On the other hand, a direct
calculation instead of the saddle point method was not sucessful either. This
is mainly because the true leading order amplitudes for states with $m\neq0$
drop from energy order $E^{4m}$ to $E^{2m}$ \cite{ChanLee1,ChanLee2, CHL} ,
and one needs to calculate the complicated subleading order contraction terms.
For this reason, the complete forms of the fixed angle closed string and
$(s,t)$ channel open string scattering amplitudes were not calculable.
However, a simple case of the $(s,t)$ channel scattering amplitude, which is
calculable for all energies, with $k_{2}$ the highest spin state $V_{2}%
=\alpha_{-1}^{\mu_{1}}\alpha_{-1}^{\mu_{2}}\cdot\cdot\alpha_{-1}^{\mu_{n}}%
\mid0,k>$ at mass level $M_{2}^{2}$ $=2(n-1)$ and three tachyons $k_{1,3,4}$
is \cite{CHL}%
\begin{equation}
A_{n}^{\mu_{1}\mu_{2}\cdot\cdot\mu_{n}}(s,t)=\overset{n}{\underset{l=0}{\sum}%
}(-)^{l}\left(  _{l}^{n}\right)  B(-\frac{s}{2}-1+l,-\frac{t}{2}%
-1+n-l)k_{1}^{(\mu_{1}}..k_{1}^{\mu_{n-l}}k_{3}^{\mu_{n-l+1}}..k_{3}^{\mu
_{n})}. \label{6}%
\end{equation}
The high energy limit of Eq.(\ref{6}) can then be calculated to be
\cite{Closed}%
\begin{equation}
A^{(n,0,0)}(s,t)=(-)^{n}\frac{\sin\left(  \pi u/2\right)  }{\sin\left(  \pi
s/2\right)  }A^{(n,0,0)}(t,u). \label{7}%
\end{equation}
The factor $\frac{\sin\left(  \pi u/2\right)  }{\sin\left(  \pi s/2\right)  }$
which was missing in the literature \cite{GM,Veneziano} has important physical
interpretations. The presence of poles give infinite number of resonances in
the string spectrum and zeros give the coherence of string scatterings. These
poles and zeros survive in the high energy limit and can not be dropped out.
Presumably, the factor triggers the failure of saddle point calculation
mentioned above.

To calculate the complete high energy closed string scattering amplitudes, one
had to rely on calculation based on the method of decoupling of zero-norm
states, or stringy Ward identities, in the spectrum. With this new input, an
infinite number of linear relations among high energy scattering amplitudes of
different string states can be derived, and the complete ratios among high
energy closed string scattering amplitudes at each fixed mass level can be
shown to be the tensor product of two sets of $(t,u)$ channel open string
ratios in eq.(\ref{3}). The complete high energy closed string and $(s,t)$
channel open string scattering amplitudes can then be obtained by
Eqs.(\ref{5}) and (\ref{7}). An explicit calculation for the lowest mass level
case was presented in \cite{Closed}. Another independent method to obtain the
closed string ratios is to calculate high energy string scattering amplitudes
in the Regge regime, which we will discuss in the next section.

\section{Regge Scattering}

Another high energy regime of string scattering amplitudes, which contains
complementary information of the theory, is the fixed momentum transfer or
Regime regime. That is in the kinematic regime%
\begin{equation}
s\rightarrow\infty,\sqrt{-t}=\text{fixed (but }\sqrt{-t}\neq\infty). \label{8}%
\end{equation}
It was found \cite{KLY} that the number of high energy scattering amplitudes
for each fixed mass level in this regime is much more numerous than that of
fixed angle regime calculated previously. On the other hand, it seems that
both the saddle-point method and the method of decoupling of zero-norm states
adopted in the calculation of fixed angle regime do not apply to the case of
Regge regime. However the calculation is still manageable, and the general
formula for the high energy $(s,t)$ channel open string scattering amplitudes
at each fixed mass level can be written down explicitly.

It was shown that the most general high energy open string states in the Regge
regime at each fixed mass level $n=\sum_{n,m}lk_{n}+mq_{m}$ are%
\begin{equation}
\left\vert k_{l},q_{m}\right\rangle =\prod_{l>0}(\alpha_{-l}^{T})^{k_{l}}%
\prod_{m>0}(\alpha_{-m}^{L})^{q_{m}}|0,k\rangle. \label{9}%
\end{equation}
For our purpose here, however, we will only calculate scattering amplitudes
corresponding to the vertex in Eq.(\ref{2}). The relevant kinematics are%
\begin{equation}
e^{P}\cdot k_{1}\simeq-\frac{s}{2M_{2}},\text{ \ }e^{P}\cdot k_{3}\simeq
-\frac{\tilde{t}}{2M_{2}}=-\frac{t-M_{2}^{2}-M_{3}^{2}}{2M_{2}}; \label{10}%
\end{equation}%
\begin{equation}
e^{L}\cdot k_{1}\simeq-\frac{s}{2M_{2}},\text{ \ }e^{L}\cdot k_{3}\simeq
-\frac{\tilde{t}^{\prime}}{2M_{2}}=-\frac{t+M_{2}^{2}-M_{3}^{2}}{2M_{2}};
\label{11}%
\end{equation}
and%
\begin{equation}
e^{T}\cdot k_{1}=0\text{, \ \ }e^{T}\cdot k_{3}\simeq-\sqrt{-{t}}. \label{12}%
\end{equation}
The Regge scattering amplitude for the $(s,t)$ channel was calculated to be
\cite{KLY}%
\begin{align}
R^{(n,2m,q)}(s,t)  &  =B\left(  -1-\frac{s}{2},-1-\frac{t}{2}\right)
\sqrt{-t}^{n-2m-2q}\left(  \frac{1}{2M_{2}}\right)  ^{2m+q}\nonumber\\
&  \cdot2^{2m}(\tilde{t}^{\prime})^{q}U\left(  -2m\,,\,\frac{t}{2}%
+2-2m\,,\,\frac{\tilde{t}^{\prime}}{2}\right)  . \label{13}%
\end{align}
In Eq.(\ref{13}) $U$ is the Kummer function of the second kind and is defined
to be%
\begin{equation}
U(a,c,x)=\frac{\pi}{\sin\pi c}\left[  \frac{M(a,c,x)}{(a-c)!(c-1)!}%
-\frac{x^{1-c}M(a+1-c,2-c,x)}{(a-1)!(1-c)!}\right]  \text{ \ }(c\neq2,3,4...)
\label{14}%
\end{equation}
where $M(a,c,x)=\sum_{j=0}^{\infty}\frac{(a)_{j}}{(c)_{j}}\frac{x^{j}}{j!}$ is
the Kummer function of the first kind. Note that the second argument of Kummer
function $c=\frac{t}{2}+2-2m,$ and is not a constant as in the usual case.

We now proceed to calculate the Regge $(t,u)$ channel scattering amplitude.
The high energy limit of the amplitude can be written as%
\begin{align}
R^{(n,2m,q)}(t,u) &  =\int_{1}^{\infty}dx\,x^{k_{1}\cdot k_{2}}(x-1)^{k_{2}%
\cdot k_{3}}\left[  \frac{e^{T}\cdot k_{3}}{1-x}\right]  ^{n-2m-2q}\nonumber\\
\cdot &  \left[  \frac{e^{L}\cdot k_{1}}{-x}+\frac{e^{L}\cdot k_{3}}%
{1-x}\right]  ^{2m}\left[  \frac{e^{L}\cdot k_{1}}{x^{2}}+\frac{e^{L}\cdot
k_{3}}{(1-x)^{2}}\right]  ^{q}\nonumber\\
&  \simeq(\sqrt{-{t}})^{n-2m-2q}\left(  \frac{\tilde{t}^{\prime}}{2M_{2}%
}\right)  ^{2m+q}\sum_{j=0}^{2m}{\binom{2m}{j}}\left(  -\right)  ^{j}\left(
\frac{s}{\tilde{t}^{\prime}}\right)  ^{j}\nonumber\\
&  \cdot(-)^{k_{2}\cdot k_{3}}\int_{1}^{\infty}dx\,x^{k_{1}\cdot k_{2}%
-j}(1-x)^{k_{2}\cdot k_{3}+j-n}.\label{15}%
\end{align}
We can make a change of variable $y=\frac{x-1}{x}$ to transform the integral
of Eq.(\ref{15}) to%
\begin{align}
R^{(n,2m,q)}(t,u) &  =(\sqrt{-{t}})^{n-2m-2q}\left(  \frac{\tilde{t}^{\prime}%
}{2M_{2}}\right)  ^{2m+q}(-)^{n}\nonumber\\
&  \cdot\sum_{j=0}^{2m}{\binom{2m}{j}}\left(  \frac{s}{\tilde{t}^{\prime}%
}\right)  ^{j}\int_{0}^{1}dy\,y^{k_{2}\cdot k_{3}+j-n}(1-y)^{n-k_{1}\cdot
k_{2}-k_{2}\cdot k_{3}-2}.\nonumber\\
&  =(\sqrt{-{t}})^{n-2m-2q}\left(  \frac{\tilde{t}^{\prime}}{2M_{2}}\right)
^{2m+q}(-)^{n}\nonumber\\
&  \cdot\sum_{j=0}^{2m}{\binom{2m}{j}}\left(  \frac{s}{\tilde{t}^{\prime}%
}\right)  ^{j}B(k_{2}\cdot k_{3}+j-n+1,n-k_{1}\cdot k_{2}-k_{2}\cdot
k_{3}-1).\label{16}%
\end{align}
In the Regge limit, the beta function can be approximated by%
\begin{align}
&  B(k_{2}\cdot k_{3}+j-n+1,n-k_{1}\cdot k_{2}-k_{2}\cdot k_{3}-1)\nonumber\\
&  =B(-1-\frac{t}{2}+j,-1-\frac{u}{2})\nonumber\\
&  \simeq B(-1-\frac{t}{2},-1-\frac{u}{2})(-1-\frac{t}{2})_{j}(\frac{s}%
{2})^{-j}\label{17}%
\end{align}
where $(a)_{j}=a(a+1)(a+2)...(a+j-1)$ is the Pochhammer symbol. In the above
calculation, we have used $s+t+u=2n-8.$ Finally, the $(t,u)$ channel amplitude
can be written as%
\begin{align}
R^{(n,2m,q)}(t,u) &  =(-)^{n}B(-1-\frac{t}{2},-1-\frac{u}{2})(\sqrt{-{t}%
})^{n-2m-2q}\left(  \frac{1}{2M_{2}}\right)  ^{2m+q}\nonumber\\
&  \cdot2^{2m}(\tilde{t}^{\prime})^{q}U\left(  -2m\,,\,\frac{t}{2}%
+2-2m\,,\,\frac{\tilde{t}^{\prime}}{2}\right)  .\label{18}%
\end{align}
We can now explicitly write down the general formula for high energy closed
string scattering amplitude corresponding to the closed string state%
\begin{equation}
\left\vert n;2m,2m^{^{\prime}};q,q^{^{\prime}}\right\rangle \equiv(\alpha
_{-1}^{T})^{\frac{n}{2}-2m-2q}(\alpha_{-1}^{L})^{2m}(\alpha_{-2}^{L}%
)^{q}\otimes(\tilde{\alpha}_{-1}^{T})^{\frac{n}{2}-2m^{^{\prime}}%
-2q^{^{\prime}}}(\tilde{\alpha}_{-1}^{L})^{2m^{^{\prime}}}(\tilde{\alpha}%
_{-2}^{L})^{q^{^{\prime}}}|0,k\rangle.\label{19}%
\end{equation}
By using Eqs.(\ref{5}), (\ref{13}) and (\ref{18}), the amplitude is%
\begin{align}
R_{\text{closed}}^{\left(  n;2m,2m^{^{\prime}};q,q^{^{\prime}}\right)
}\left(  s,t,u\right)   &  =(-)^{n}\sin\left(  \pi k_{2}\cdot k_{3}\right)
B(-1-\frac{s}{2},-1-\frac{t}{2})B(-1-\frac{t}{2},-1-\frac{u}{2})\nonumber\\
&  \cdot(\sqrt{-{t}})^{n-2(m+m^{^{\prime}})-2(q+q^{^{\prime}})}\left(
\frac{\tilde{t}^{\prime}}{2M_{2}}\right)  ^{2(m+m^{^{\prime}})+q+q^{^{\prime}%
}}2^{(2m+m^{^{\prime}})}(\tilde{t}^{\prime})^{q+q^{^{\prime}}}\nonumber\\
&  \cdot U\left(  -2m\,,\,\frac{t}{2}+2-2m\,,\,\frac{\tilde{t}^{\prime}}%
{2}\right)  U\left(  -2m^{^{\prime}}\,,\,\frac{t}{2}+2-2m^{^{\prime}%
}\,,\,\frac{\tilde{t}^{\prime}}{2}\right)  .\label{20}%
\end{align}
The Regge scattering amplitudes at each fixed mass level are no longer
proportional to each other. The ratios are $t$ dependent functions and can be
calculated to be%
\begin{align}
\frac{R^{(n,2m,q)}(s,t)}{R^{(n,0,0)}(s,t)} &  =(-1)^{m}\left(  -\frac
{1}{2M_{2}}\right)  ^{2m+q}(\tilde{t}^{\prime}-2N)^{-m-q}(\tilde{t}^{\prime
})^{2m+q}\nonumber\\
\cdot &  \sum_{j=0}^{2m}(-2m)_{j}\left(  -1+n-\frac{\tilde{t}^{\prime}}%
{2}\right)  _{j}\frac{(-2/\tilde{t}^{\prime})^{j}}{j!}+\mathit{O}\left\{
\left(  \frac{1}{t}\right)  ^{m+1}\right\}  .\label{21}%
\end{align}
An interesting observation \cite{KLY} is that the coefficients of the leading
power of $\tilde{t}^{\prime}$ in Eq. (\ref{21}) can be identified with the
ratios in Eqs.(\ref{3}). To ensure this identification, we need the following
identity
\begin{align}
&  \sum_{j=0}^{2m}(-2m)_{j}\left(  -1+n-\frac{\tilde{t}^{\prime}}{2}\right)
_{j}\frac{(-2/\tilde{t}^{\prime})^{j}}{j!}\nonumber\\
&  =0(-\tilde{t}^{\prime})^{0}+0(-\tilde{t}^{\prime})^{-1}+...+0(-\tilde
{t}^{\prime})^{-m+1}+\frac{(2m)!}{m!}(-\tilde{t}^{\prime})^{-m}+\mathit{O}%
\left\{  \left(  \frac{1}{\tilde{t}^{\prime}}\right)  ^{m+1}\right\}
.\label{22}%
\end{align}
Note that $n$ effects only the sub-leading terms in $\mathit{O}\left\{
\left(  \frac{1}{\tilde{t}^{\prime}}\right)  ^{m+1}\right\}  .$ Eq.(\ref{21})
was exactly proved \cite{KLY} for $n=0,1$ by using Stirling number identities
developed in combinatoric number theory \cite{MK}. For general integer $n$
case, only the identity corresponging to the term $\frac{(2m)!}{m!}(-\tilde
{t}^{\prime})^{-m}$ was rigoursly proved \cite{HLTY} but not other "0
identities". We conjecture that Eq. (\ref{22}) is valid for any \textit{real}
number $n.$ We have numerically shown the validity of Eq. (\ref{22}) for the
value of $m$ up to $m=10.$ Here we give only results of $m=3$ and $4$
\begin{align}
&  \sum_{j=0}^{6}(-2m)_{j}\left(  -1+n-\frac{\tilde{t}^{\prime}}{2}\right)
_{j}\frac{(-2/\tilde{t}^{\prime})^{j}}{j!}\nonumber\\
&  =\frac{120}{(-\tilde{t}^{\prime})^{3}}+\frac{720a^{2}+2640a+2080}%
{(-\tilde{t}^{\prime})^{4}}+\frac{480a^{4}+4160a^{3}+12000a^{2}+12928a+3840}%
{(-\tilde{t}^{\prime})^{5}}\nonumber\\
&  +\frac{64a^{6}+960a^{5}+5440a^{4}+14400a^{3}+17536a^{2}+7680a}{(-\tilde
{t}^{\prime})^{6}},\label{23}%
\end{align}%
\begin{align}
&  \sum_{j=0}^{8}(-2m)_{j}\left(  -1+n-\frac{\tilde{t}^{\prime}}{2}\right)
_{j}\frac{(-2/\tilde{t}^{\prime})^{j}}{j!}\nonumber\\
&  =\frac{1680}{(-\tilde{t}^{\prime})^{4}}+\frac{13440a^{2}+67200a+76160}%
{(-\tilde{t}^{\prime})^{5}}\nonumber\\
&  +\frac{13440a^{4}+152320a^{3}+595840a^{2}+930048a+467712}{(-\tilde
{t}^{\prime})^{6}}\nonumber\\
&  +\frac{3584a^{6}+68096a^{5}+501760a^{4}+1802752a^{3}+3236352a^{2}%
+2608128a+645120}{(-\tilde{t}^{\prime})^{7}}\nonumber\\
&  +\frac{256a^{8}+7168a^{7}+82432a^{6}+501760a^{5}+1732864a^{4}%
+3361792a^{3}+3345408a^{2}+1290240a}{(-\tilde{t}^{\prime})^{8}}\label{24}%
\end{align}
where $a=-1+n.$ We can see that $a$ shows up only in the sub-leading order
terms as expected. From the form of Eq.(\ref{20}), we conclude that the high
energy closed string ratios in the fixed angle regime can be extracted from
Kummer functions and are calculated to be
\begin{align}
\frac{A_{\text{closed}}^{\left(  n;2m,2m^{^{\prime}};q,q^{^{\prime}}\right)
}\left(  s,t,u\right)  }{A_{\text{closed}}^{\left(  n;0,0;0,0\right)  }\left(
s,t,u\right)  } &  =\left(  -\frac{1}{M_{2}}\right)  ^{2(m+m^{^{\prime}%
})+q+q^{^{\prime}}}\left(  \frac{1}{2}\right)  ^{q+q^{^{\prime}}}\nonumber\\
&  \lim_{t\rightarrow\infty}(-t)^{-m-m^{^{\prime}}}U\left(  -2m\,,\,\frac
{t}{2}+2-2m\,,\,\frac{t}{2}\right)  U\left(  -2m^{^{\prime}}\,,\,\frac{t}%
{2}+2-2m^{^{\prime}}\,,\,\frac{t}{2}\right)  \nonumber\\
= &  \left(  -\frac{1}{M_{2}}\right)  ^{2(m+m^{^{\prime}})+q+q^{^{\prime}}%
}\left(  \frac{1}{2}\right)  ^{m+m^{^{\prime}}+q+q^{^{\prime}}}%
(2m-1)!!(2m^{^{\prime}}-1)!!.\label{25}%
\end{align}
This is an alternative method to calculate the high energy closed string
ratios other than the method of decoupling of zero norm state\ adopted
previously. In addition to redriving the ratios calculated previously, one can
express the ratios in terms of Kummer functions through the Regge calculation
presented in this paper. This may turn out to be important for the
understanding of algebraic structure of stringy symmetry.

In conclusion, a direct calculation of general formula for high energy closed
string scattering amplitudes is doable in the Regge regime and is calculated
in Eq.(\ref{20}), but not in the fixed angle regime. The ratios among high
energy closed string scattering amplitudes for each fixed mass level in the
fixed angle regime, which were calculated previously by the method of
decoupling of zero norm states, can be alternatively deduced from general
formula of high energy closed string scattering amplitudes in the Regge
regime.\noindent\ The result that the ratios can be expressed in terms of
Kummer functions in the Regge calculation presented in this paper may help to
understand the algebraic structure of stringy symmetry.

\section{Acknowledgments}

We thank Rong-Shing Chang, Song He, Yoshihiro Mitsuka and Keijiro Takahashi
for helpful discussions. This work is supported in part by the National
Science Council, 50 billions project of Ministry of Education and National
Center for Theoretical Science, Taiwan.

\end{document}